\documentclass[11pt,twoside]{article}

\pdfoutput=1

\usepackage{asp2006}
\usepackage{lscape}
\usepackage{graphicx}

\begin{document}

\title{Measuring Solar Abundances with Seismology}

\author{Katie Mussack}
  \affil{Institute of Astronomy, University of Cambridge, Cambridge, CB3 0HA,
  UK} 

\author{Douglas Gough}
  \affil{Institute of Astronomy, University of Cambridge, Cambridge, CB3 0HA,
  UK
  and 
  Department of Applied Mathematics \& Theoretical Physics,
  Wilberforce Road, Cambridge, CB3 0WA, UK}

\begin{abstract}
The revision of the photospheric
abundances proferred by Asplund et al has rendered opacity theory inconsistent
with the seismologically determined opacity through the Sun. This highlights
the need for a direct seismological measurement of solar abundances. Here we
describe the technique used to measure abundances with seismology, examine our
ability to detect differences between solar models using this technique, and
discuss its application in the Sun.

\end{abstract}

\section{Introduction}

A decade ago, solar models matched the Sun remarkably well. Sound speed
profiles in models differed from the Sun's actual sound speed by less than
half a percent. Then \citet{KAM_Asplund_2005} reanalyzed the solar optical
spectrum using improved atomic physics and a 3D hydrodynamical model of the
atmosphere (instead of a 1D hydrostatic model). Their results indicate that
the abundances of the heavy elements should be lower by a substantial
amount. For example, carbon was lowered by 35\%, nitrogen by 27.5\%, oxygen by
48\%, and neon by 74\%.  Solar models that are evolved using the new abundance
mixture give worse agreement with helioseismic constraints than models using
the old abundances. Problems include a tripling in the sound-speed discrepency
below the convection zone, a convection zone that is too shallow, and a
helium abundance in the convection zone that is too low. 

Many attempts have
been made to reconcile the new abundances with seismology. A few examples
include increasing the opacities below the convection zone, increasing
abundances within their uncertainties, increasing the neon abundance,
enhancing diffusive settling rates, and including early accretion of lower-Z
material. Various combinations of these changes have also been explored. None
of the alterations has resolved the discrepency between the revised
abundances and seismology. For further discussion of the problem and the
attempted solutions, see
\citet{KAM_Basu_2008} and \citet{KAM_Guzik_2008} in these proceedings. Since the
mismatch between seismology and the new abundances remains a problem, a
seismic determination of the heavy element abundances in the solar convection
zone is necessary. The techniques for a seismic measurement of heavy elements
are discussed in the following sections.

\section{Techniques}

\begin{figure}[!ht]
 \centering
 \includegraphics[scale=0.8]{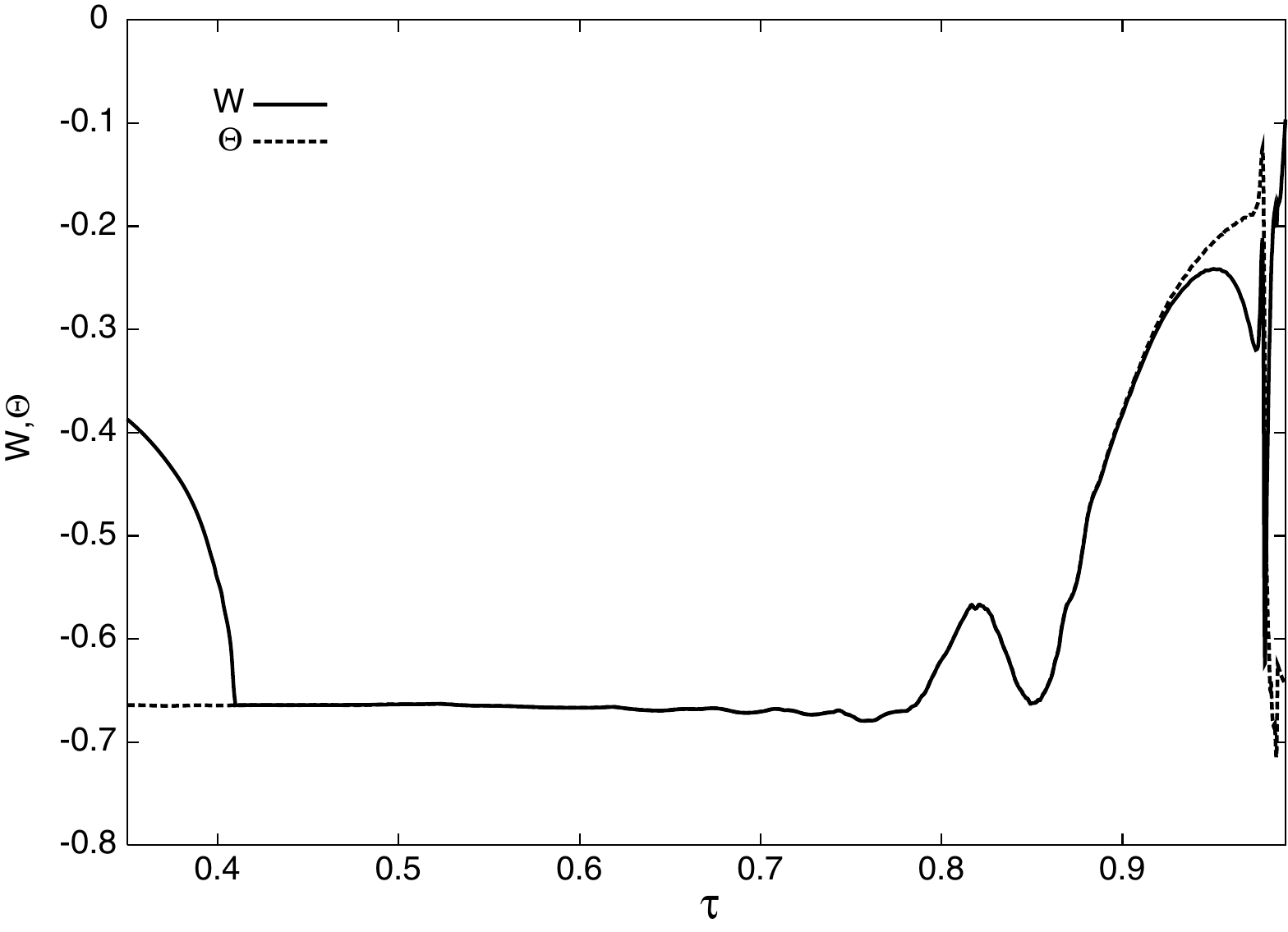}
 \caption{$W$ and $\Theta$ throughout the convection zone in model S of
 \citet{KAM_JCD_1996}. The large hump
 around $\tau=0.82$ is due to helium ionization. The smaller wiggles lower in
 the convection zone are due to heavy element ionization.}
 \label{figone}
\end{figure}

\citet{KAM_Dappen_1984} measured the solar helium abundance through the effect
of helium
ionization on the adiabatic exponent $\gamma_1$. Ionization lowers
$\gamma_1$ from the ideal gas value of 5/3. Since the convection zone
stratification is almost purely adiabatic, the gradient of the square of the
sound speed in that region can be written as
\begin{equation}
 \label{eq1}
 \frac{\textrm{d}c^2}{\textrm{d}r}=\frac{Gm}{r^2} \left\{ 1-\gamma_1
 \left[1+\left(\frac{\partial \ln{\gamma_1}}{\partial \ln p}\right)_s \right]
 \right\}. 
\end{equation}
This expression can be separated into a term called $W$
that contains the seismic variables and a term called $\Theta$ that contains
only thermodynamic variables.
\begin{equation}
\label{eq2}
 W = \frac{r^2}{Gm} \frac{\textrm{d}c^2}{\textrm{d}r}
\end{equation}
\begin{equation} 
 \label{eq3}
 \Theta = 1 - \gamma_1 \left[ 1+\left( \frac{\partial \ln \gamma_1}{\partial
 \ln p}\right)_s \right] 
\end{equation}
Since $\Theta$ contains a derivative of
$\gamma_1$, it shows a more pronounced variation due to the ionization of
each element than
$\gamma_1$ alone shows. For the helium abundance measurement, D\"appen
and Gough used seismic data to obtain the sound speed, then computed $W$ and
$\Theta$ for their model. Figure \ref{figone} shows the helium hump
that D\"appen and Gough discovered in $W$ and $\Theta$ where \ion{He}{II}
ionization takes place in the convection zone. Here $W$ and $\Theta$ are
ploted against the acoustic radius $\tau=\int c^{-1} \textrm{d}r$. The
signatures of heavy-element ionization can already be seen in the
small wiggles lower in the convection zone \citep{KAM_Gough_2006}. Near
$\tau=0.63$, we see \ion{C}{V} 
ionization. From 0.45-0.55, \ion{O}{VII} ionization appears. Near 0.58,
\ion{N}{VI} shows up as an imperceptibly small hump that is impossible to
measure here. Near 0.68, there is a double hump due to the first two K-shell
ionizations of Ne. The larger humps are merged ionization zones. Since the
humps from heavy element ionization are such small features,
great care must be taken to obtain an accurate measurement of the abundances.

\subsection{Inversion}

It is
common to use sound speed and density or sound speed and $\gamma_1$ as solar
inversion variables. However, the extremely small ionization humps of the
heavy elements require a more sensitive inversion. We suggest inverting the
frequency differences directly for $W$ and $\Theta$. 
\begin{equation}
 \frac{\delta \omega^2}{\omega^2} = \int_0^R{\left(K_{W,\Theta} \delta W +
K_{\Theta, W} \delta \Theta \right) \textrm{d}r}
\end{equation}
In the adiabatically
stratified convection zone, $W = \Theta$. This allows us to rewrite the
integral as 
\begin{equation}
 \frac{\delta \omega^2}{\omega^2} = \int_0^R{\left(K_{W,\Theta} + K_{\Theta, W}
 \right) \delta \Theta \: \textrm{d}r} .
\end{equation}
Since this formulation has the added advantage of reducing the problem
to a single-variable inversion, we no longer need to worry about contamination
from a second inversion variable.

\subsection{Isothermal Sound Speed}

\begin{figure}[!ht]
 \centering
 \includegraphics[scale=0.8]{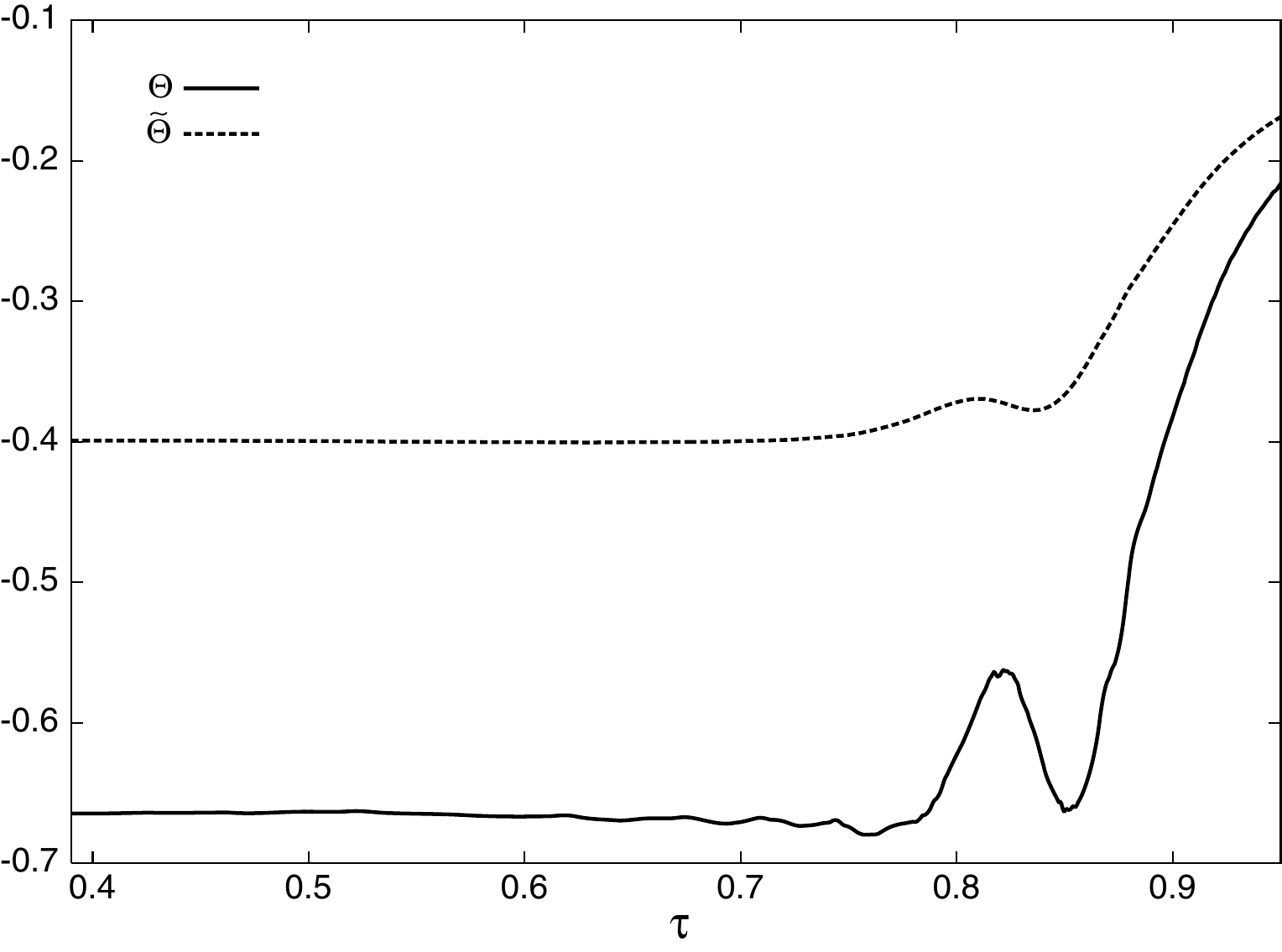}
 \caption{
  $\Theta$ and  $\tilde{\Theta}$ 
 throughout the convection zone in model S of \citet{KAM_JCD_1996}.  The
 heavy-element ionization humps are clearly visible in $\Theta$, yet they do
 not appear in $\tilde{\Theta}$.
 }
 \label{figtwo}
\end{figure}

The conversion from the familiar sound speed and density kernels \citep[see,
for example,][]{KAM_Gough_1991} to the desired $W$ and $\Theta$ kernels requires
quite a bit of 
algebra. One possible trick to simplify the calculation is to
reconfigure $W$ and $\Theta$ in terms of the square of the isothermal
sound speed $u = p/\rho$. The gradient of $u$ in an adiabatically
stratified region is 
\begin{equation}
 \frac{\textrm{d}u}{\textrm{d}r}= \frac{Gm}{r^2}  \left( \frac{1}{\gamma_1}
 -1\right) . 
\end{equation}
This leads to a set of definitions analogous to equations
(\ref{eq2}) and (\ref{eq3}) for the separation of seismic and thermodynamic
variables:
\begin{equation}
 \tilde{W} = \frac{r^2}{Gm} \frac{\textrm{d}u}{\textrm{d}r},
\end{equation}
\begin{equation}
 \tilde{\Theta} =\frac{1}{\gamma_1} -1 .
\end{equation}
The derivative of $\gamma_1$ in the
definition of $\Theta$ does not appear in the new $\tilde{\Theta}$. This
greatly simplifies the kernel conversion. However, without the $\gamma_1$
derivative, $\tilde{\Theta}$ is less sensitive to the effects of ionization
than $\Theta$ is. As shown in figure \ref{figtwo}, $\tilde{\Theta}$ does not
have visible heavy-element ionization humps. Since the effect of ionization is
so small to begin with, reducing it even further by switching to
$\tilde{\Theta}$ is not worth the gain in algebraic simplicity in the kernel
conversion. The $\gamma_1$ derivative is essential to highlight the effect of
heavy-element ionization. We therefore choose to use the original definitions
of $W$ and $\Theta$.

\subsection{Pressure Ionization}

One source of
uncertainty in solar equations of state is the effect of pressure
ionization. Pressure ionization occurs when matter is so dense that electron
orbits overlap.  The electrons in the overlapping orbits no longer belong to a
particular atomic nucleus.  The pressure effectively frees these electrons,
increasing the ionization of the plasma. \citet{KAM_Baturin_2000}
developed a technique to account for pressure ionization in an equation of
state by applying constraints to the effective atomic and ionic sizes. They
used a single pressure-ionization parameter $\zeta$ for all species of
ions. They then implemented a model in which they calibrated $\zeta$ to fit
solar data. The calibrated pressure-ionization model fit the data better than
the parameter-free equations of state, highlighting the importance of pressure
ionization for an accurate equation of state.

Changing the effective atomic size in the pressure-ionization model leads to a
vertical shift in $\gamma_1$. Since $\gamma_1$ appears as a factor in $\Theta$,
$\Theta$ also suffers from this shift due to the atomic size
parameterization. Therefore, we cannot just measure the average value of
$\Theta$ to obtain the heavy-element abundances. In order to obtain an
accurate seismological measurement, we must consider the effective atomic size
by using a pressure-ionization model for the equation of state or measure the
humps relative to the background level of $\Theta$. 

\section{Conclusions}

In order to obtain a seismic measurement
of the heavy-element abundances in the Sun the following techniques should be
used: the frequency data should be inverted directly for $W$ or $\Theta$
instead of sound speed, the original definitions for $W$ and $\Theta$
should be used instead of the $\tilde{W}$ and $\tilde{\Theta}$ derived from
the isothermal sound speed formulation, and the
effect of atomic size calibration on the background level of $\Theta$ should
be taken into account. These choices are crucial for accurately measuring the
small effect of heavy-element ionization in order to determine solar
abundances seismically.

\end{document}